\documentclass[preprint2]{aastex}
\usepackage{graphicx,natbib}
\begin{document}

\title{\bf  Scale-Free Magnetic Networks:  Comparing Observational Data with
a Self-Organizing Model of the Coronal Field}

\author{David Hughes and Maya Paczuski}\affil{Mathematical
Physics, Department of Mathematics, Huxley Building, Imperial College of
Science, Technology and Medicine, London UK SW7 2BZ}
\email{maya@imperial.ac.uk}

\begin{abstract}
We propose that the coronal magnetic field, linking concentrations on the
photosphere through an interwoven web of flux, embodies a scale-free network.
It arises from a self-organized critical dynamics including flux emergence, the
diffusion and merging of magnetic concentrations, as well as avalanches of
reconnecting flux tubes. Magnetic concentrations such as fragments, pores and
sunspots, are `nodes' joined by flux tubes or `links'.  The number of links
emanating from a node is scale-free.  We reanalyze the quiet-Sun data of Close
et al and show that the distribution of magnetic concentration strengths is a
power law with an index $\gamma= 1.7 \pm 0.3$, over the entire range of the
measurement, about $(2-500)\times 10^{17}$ Mx. This distribution is compatible
with that for the sizes of active regions reported by Harvey and Schwaan. Thus
magnetic concentrations may be scale-free from the smallest measurable
fragments to the large active regions. Numerical simulations of a
self-organized critical model give the same index $\gamma$, within statistical
uncertainty. The exponential distribution of flux tube lengths also agrees
quantitatively with results from the model, below the supergranule cell size.
Calibration with the measured diffusion constant of magnetic concentrations
allows us to calculate a flux turnover time in the model to be of order 10
hours and the total solar flux to be of order $10^{23}$Mx, agreeing with
observations. We introduce two other statistical quantities to characterize
scale-free networks. The probability distribution for the amount of flux
connecting a pair of concentrations, and the number of distinct concentrations
linked to a given one are predicted to be scale-free, with different indices.
Our approach unifies the observation of scale free flare energies with the
coronal magnetic field structure.
\end{abstract}

\keywords{ Sun:corona --- Sun:magnetic fields --- Sun: photosphere}


\section{Introduction}

A complex interwoven network of magnetic fields threads the surface of the sun.
Magnetic energy stored in the coronal network builds up due to turbulent plasma
forces below the photosphere until stresses are suddenly released by
reconnection \citep{mhd, parkerbook}.  If the magnetic energy released is
sufficiently large, it induces radiative emission that is detected as a flare
event. Flares are often classified into hard X-ray flares, EUV transient
brightenings, nanoflares, etc. However, observations show that one cannot
associate any length, time, or energy scale to flares overall. In particular,
the probability distribution of flare energies appears as a featureless power
law that spans more than eight orders of magnitude, which is the entire
observable range \citep{aschwanden:TRACE}. This points to a scale-invariant
energy release mechanism.  \citet{lh} first proposed that the corona is in a
self-organized critical (SOC) state, with avalanches of all sizes \citep{btw}.
In this picture, all flares large and small, come from avalanches of magnetic
reconnection.  SOC systems often show scale free behavior not only for their
event statistics but also for emergent spatial and temporal structures, as seen
both in physical systems (e.g. \citet{frette:realrice}) and numerical models
(e.g. \citet{hughes:nsdm}). For a review see \citet{soc}.

Like flares, concentrations of magnetic flux on the photosphere also exist on a
wide variety of scales. The strongest concentrations are sunspots, which occur
in active regions that may contain more than $10^{22}$Mx, at smaller scales are
ephemeral regions, pores and, at the smallest resolvable concentrations above
the current resolution scale of $\approx 10^{16}$Mx, are fragments.  The
structure of the quiet-Sun has been described as a `magnetic carpet' with
magnetic concentrations of various strengths linked by flux tubes
\citep{schrijver_flux}.

Together with Dendy, Helander, and McClements, we recently introduced a
multi-loop model of the coronal magnetic field that exhibits self-organized
criticality \citep{modelprl}.  Photospheric magnetic flux concentrations,
represented by points on a surface, are connected by loops, representing flux
tubes. The model exhibits reconnection avalanches of all sizes, and thus a
power law distribution of flare events. The structure of the magnetic network
in the model qualitatively resembles the magnetic carpet, as shown in Figure 1.
Its structure is mathematically described as a scale-free network
\citep{barabasi99}. The model provides a single physical process which gives
rise to both a scale free magnetic field structure as well as a power law
distribution of flares. Thus flare event statistics are unified with the scale
free geometry of the coronal magnetic field.

\begin{figure}[t]
\plotone{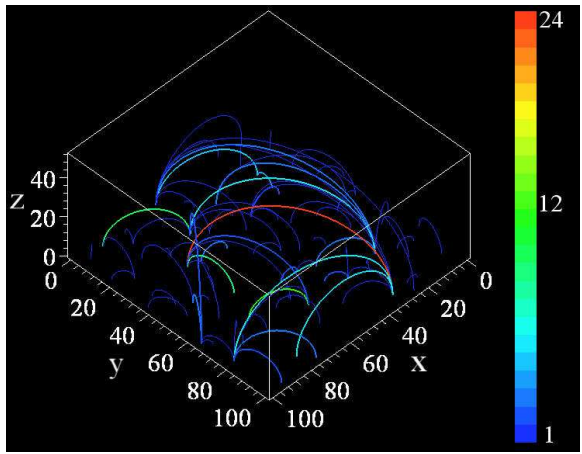} \caption{Snapshot of loops in the steady state of the model.
Footpoints lie in the $(xy)$ plane and are linked by loops. The loops are
colored to indicate the relative strength of the connection, as shown by the
scale to the right. Note that there is a large range in the number of loops
emanating from different footpoints, as well as a wide range of connection
strengths. The figure was generated with model I using parameters $L=100$,
$m=1$.} \label{loops}
\end{figure}

Here, we make a detailed study of the properties of the network. This provides
a quantitative basis enabling direct statistical comparison between the network
structure of the coronal magnetic field and of the model. To this end, we
reanalyze recent measurements of the distribution of strengths of magnetic flux
from \citet{close} and \citet{hagenaar2003}. We find that the strength of
magnetic flux concentrations are scale-free, as predicted by the model.

Based on this quantitative statistical comparison, we propose that the coronal
magnetic field embodies a scale-free network that emerges as a consequence of a
SOC dynamics involving avalanches of reconnection. Magnetic concentrations on
the photosphere are nodes in the network while flux tubes are links connecting
the nodes. The defining signature of a scale-free network is that the number of
links attached to any node is distributed according to a power law. Physically,
this means that the amount of flux emanating from a magnetic concentration on
the photosphere is scale-invariant up to a cutoff determined by the overall
system size -- in this case the total amount of flux emanating from the
photosphere.

Scale-free networks arise in many different contexts, including the internet
\citep{faloustos}, the World Wide Web \citep{barabasi99}, interaction and
regulatory networks \citep{wagner01,maslov02}, or the citation network
\citep{redner:citations}, for a review see e.g. \citet{barabasi02, bornholdt}.
There is currently an extensive research effort to study the consequences of
that observation in biological and social phenomena. Unlike previous models, in
our model the network itself emerges as a consequence of the dynamics, and is
not, a priori, forced or assumed to exist as a system-wide network.

Section 2 gives a general explanation of the model and then a detailed
definition.  The latter may be skipped by readers primarily interested in the
results.  Section 3 contains  a quantitative comparison between numerical
simulation results and reanalyzed observational data for the distribution of
magnetic concentration strengths. A single calibration is used, setting the
loop unit in the model equal to the minimum magnetic flux threshold that can
currently be detected. The connection with scale-free networks is established.
In Section 4 we compare the exponential distribution of flux tube lengths in
the model with observational data of Close et al; this allows a calibration of
the length unit in the model to distances on the photosphere. This length scale
calibration allows a further calibration of the time unit of the model by using
measurements of the diffusion constant of magnetic concentrations
\citep{hagenaar99}. We thereby calculate the flux emergence and turnover rate
predicted by the model, and compare with current estimates for the solar
magnetic field. We also propose new statistical quantities that can be measured
to quantify the statistical properties of the coronal network structure, and
make predictions for them. These are the distribution of the flux strengths
between pairs of opposite polarity concentrations, and the distribution of the
number of distinct concentrations connected to any given one.  These describe a
network where typical magnetic concentrations are connected to many other ones,
but most of the flux emanating from one concentration is typically linked to
only one other concentration. The main results and conclusions are summarized
in section 5.

\section{The SOC Model}

The coronal magnetic field consists of magnetic flux concentrations on the
photosphere linked by magnetic fields, or flux tubes.  Five processes have been
identified for the quiet-Sun: (i) {\it emergence} - magnetic flux tubes are
injected from beneath the photosphere, or may be removed (submergence); (ii)
{\it diffusion} - the footpoints of the flux tubes are continuously agitated by
the turbulent convection of the plasma; (iii) {\it coalescence} - footpoints of
the same sign merge to form concentrations and these concentrations can
aggregate to form even larger concentrations; (iv) {\it cancellation} - mutual
loss of flux from opposite polarity concentrations; (v) {\it fragmentation} -
the break up of concentrations into smaller ones.

All these processes refer explicitly to dynamics that can be observed in
magnetic images of the photosphere. To this we add another process, (vi) {\it
reconnection} - flux tubes linking the concentrations reconnect in the corona
when magnetic field gradients become sufficiently steep.  Reconnection of flux
tubes lowers the magnetic field energy but, by itself, does not alter strength
of the concentrations. Currently, the model does not include an explicit
fragmentation process.

\subsection{Explanation of the Model}
Although it may seem unwieldy,  it is actually straightforward to construct a
physically motivated model where these processes are unavoidably realized.  Of
course, in order to describe magnetic field evolution both at large energy and
length scales, and at the extremely high magnetic Reynolds number relevant to
the corona, it is necessary to make simplifications of the underlying physical
laws. Nevertheless, the model does retain certain physical features that are
known to be important. Primarily, it keeps track of the geometrical constraints
that the high conductivity of the corona and the physics of reconnection impose
on magnetic field evolution.  Magnetic flux is frozen into the plasma and is
constrained to move with it. However, when magnetic field gradients are steep
reconnection can occur, changing the geometry of the magnetic field structure.
Secondly, diffusion of footpoints and flux emergence are considered to be the
dominant sources driving coronal magnetic field energy.

The fundamental entity in the model is a directed loop which traces the midline
of a flux tube and is anchored to a surface at two opposite polarity
footpoints.  A collection of these loops and their footpoints gives a
representation of a coronal magnetic field structure.  It is able to describe
fields that are very complicated or interwoven, like the magnetic carpet.  A
snapshot of a configuration in the steady-state is shown in Fig.  \ref{loops}.
The number of loops connecting any pair of footpoints is indicated by a
color-coding. It is evident that both the number of loops attached to a
footpoint and the strength of those loops vary over a broad range.

Loops injected at small length scales are stretched and shrunk as their
footpoints diffuse over the surface. Loops submerge when their footpoints
approach closely. Nearby footpoints of the same polarity coalesce, to form
magnetic fragments, which can themselves coalesce to form ever larger
concentrations of flux, such as pores and sunspots. Conversely, opposite
polarity footpoints may cancel (though see section 2.3). Loops can reconnect
when they collide in three dimensional space, thereby releasing magnetic
energy. Reconnection of flux tubes or concentration cancelation may trigger a
cascade of further reconnection, representing a flare.

\subsection{Definition of the Model}

1.\ \ {\it Flux Tube:}  Each flux tube is represented by an infinitesimally
thin, directed loop which traces its midline.  A coronal magnetic field is
described by a configuration of many of these loops.  For simplicity each loop
is considered to be a semicircle emerging perpendicular to the $(xy)$ plane.
Every loop has a positive footpoint, where magnetic flux emerges from the
photosphere, and a negative one, where flux returns. The size of the system in
the $(xy)$ plane, which represents a region of the photospheric surface, is $L
\times L$.  Loops are labeled by an integer, $n$, and the positions of the two
footpoints of the $n$th loop are labeled in the $(xy)$ plane by ${\bf r}^+_n$
and ${\bf r}^-_n$.  The footpoint separation of a flux tube is then $d_n= |{\bf
r}^+_n -{\bf r}^-_n|$ and the length of a flux tube is $l_n= {\pi\over 2} |{\bf
r}^+_n -{\bf r}^-_n|$

2.\ \ {\it Magnetic Concentration:} A footpoint labels the center of a magnetic
concentration.  Due to coalescence, described in step 6, a footpoint can have
more than one loop attached to it. This means that the magnetic flux from any
concentration, or the flux connecting any two concentrations, can be
arbitrarily large, despite the fact that the individual loops, defined in step
1 represent small, quantized units of flux.  We do not attach a surface area to
the concentrations, just as we do not attach a width to flux tubes.  A
concentration in the model is completely described by a single footpoint
located at its center and the number of loops connected to the footpoint
represents the total flux of the fragment.  To compare with observations, one
loop in the model should be set equal to the minimum threshold for flux to be
included in the data set.

3.\ \ {\it Footpoint stirring:} This and the next term represent the driving
terms for magnetic flux.  The diffusion of footpoints, and therefore the
concentrations they represent, is described as a random walk on the two
dimensional $(xy)$ plane. At an update step, an arbitrary footpoint is chosen
at random and its position is moved, ${\bf r} \rightarrow {\bf r} + \Delta {\bf
r}$. The vector $\Delta{\bf r}$ has length and angle chosen randomly from
uniform distributions between 0 and $1$, and 0 and $2\pi$, respectively. If the
initial loop lengths are small, footpoint diffusion will cause an average
increase in the length of the loops. In this way, photospheric turbulence pumps
magnetic energy into the coronal field. For simplicity, we usually study a
system with open reflective boundary conditions; if a footpoint attempts to
move outside the $L \times L$ box in the $(xy)$ plane, it is elastically
reflected back into the box.

4.\ \ {\it Flux emergence and submergence:}  Small loops are injected into the
system at random locations, with footpoints initially separated by a distance
$l_{nl}=4$. Loops with footpoints closer than distance $l_{min}=2$ are removed
from the system, corresponding to loop submergence.  The precise (small) length
scales of these two processes do not effect the large scale properties of the
system. The essential feature is that at small length scales the model
dynamically maintains a flow of loops. Thus the magnetic field of the corona is
represented as an open system, and the flux in the system will turnover in time
or be replaced with new flux.

New flux is injected at a rate determined by a control parameter $m$. This
quantity is defined as follows: the number of footpoint updates (step 3) that
separate an injection of flux (step 4) is $m$ times the number of footpoints in
the system at that time. Thus $m$ is the average number of random walk steps a
footpoint will experience between new loops injected into the system.

\begin{figure}[t]
\plotone{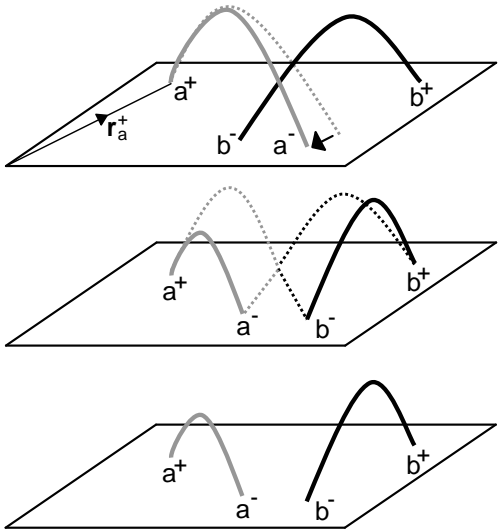} \caption{Diagram showing the process of a reconnection event,
from top to bottom. In frame 1 loop $a$ moves from its previous position
(dashed line) and crosses loop $b$. In frame 2 the loops exchange footpoints
and move to their rewired state. Frame 3 shows the final relaxed
configuration.} \label{reconnect}
\end{figure}

5. \ \ {\it Reconnection of Flux Tubes:}  Reconnection can occur when either
(a) two loops collide in three dimensional space, or (b) two footpoints cancel
or annihilate as explained in step 7.  In the first case, the midlines of two
flux tubes have crossed resulting in a strong magnetic field gradient at that
point. The flux emerging from the positive footpoint of one of the reconnecting
loops is then no longer constrained to end up at the other footpoint of the
same loop, but may instead go to the negative footpoint of the other loop (see
Figure \ref{reconnect}).  Reconnection is only allowed if it shortens the
combined length of the two colliding loops.  This process is rapid compared to
all other processes because information is transmitted along the flux tubes at
the Alfv\'{e}n speed. It occurs instantaneously in the model.

If rewiring occurs, it may happen that one or both  loops need to cross some
other loop in order to reach its rewired state. Thus a single reconnection
between a pair of loops can trigger a cascade of causally related reconnection
events. The reconnection dynamics of multiply connected footpoints that were
used in the numerical simulations is a straightforward extension (D. Hughes, in
preparation).

6. \ \ {\it Coalescence:}  If a footpoint moves within the distance $l_{min}$
of another footpoint of the same polarity, it is reassigned to the latter
footpoint's position and they move as one footpoint thereafter. The two linked
footpoints therefore form a single magnetic concentration whose center is
located at the footpoint's position.

7. \ \ {\it Cancelation:}  Footpoints of opposite polarity, belonging to
different flux tubes, can annihilate when they approach on the photosphere
\citep{livi1985,mhd}.  When footpoints of opposite polarity, belonging to
different loops, approach within a distance $l_{min}$ in the model, both
footpoints are eliminated and the remaining two footpoints are attached,
forming one loop, where before there were two.  The annihilation dynamics of
concentrations  with more than one attached loop can be implemented in
different ways (D. Hughes, in preparation). The scaling behavior of the model
does not appear to be sensitive to the precise algorithm. Footpoint
annihilation may cause collisions between loops leading to further
reconnection, as per step 5.

\subsection{Simulations and parameters}

Here we present results from numerical simulations of two different models:
model I includes cancelation (step 7) and model II does not. Both models I and
II include the submergence of small loops (as per step 4). Note that in
\citet{modelprl}, numerical results were presented only for model I.

For each version, the configuration of loops slowly evolves in response to the
driving, coalescence and reconnection processes. It reaches a dynamic steady
state whose statistical character is independent of the initial conditions.  In
this and the following sections, numerical simulation results are presented for
a range of system sizes $L=25$ to $200$ and stirring rates $m=0.001$ to $10$,
including approximately $10^7$ avalanches of reconnection events in the steady
state, for each of the two models.  The statistical behavior is robust on
varying the stirring rate $m$,  and system size $L$, even though the total
number of loops and footpoints in the system vary widely.

\section{Distribution of Magnetic Concentration Strengths}

In this section, we reanalyze observational data for the distribution of
magnetic concentration strengths, and compare with results from numerical
simulations of the SOC model.   This provides a calibration between the flux
loop in the model and magnetic flux on the photosphere. The connection with
scale-free networks is established.

\subsection{Observational data of Close et al}
\citet{close} report the cumulative distribution of fragment sizes, measured in
terms of their total magnetic flux, in a balanced section of the quiet-Sun.
They analyzed high-resolution MDI magnetograms.  The flux in a magnetogram
image of the photosphere of size 88Mm$\times$88Mm was represented by a series
of point sources.  A macropixel contained a point source if the absolute flux
density in the pixel exceeded a threshold of $1.55 \times 10^{17}$Mx. Magnetic
field lines corresponding to the potential field were traced between these
point sources. Each field line represents the same amount of magnetic flux.
Close et al argued that it is reasonable to consider the field lines as flux
tubes. Loops in our model thus correspond to the field lines calculated by
Close et al. Obviously, at the scale of the system size the net flux in our
model is zero so we are only comparing with their balanced data.

\subsubsection{A reanalysis of the data of Close et al}

\begin{figure}[t]
\plotone{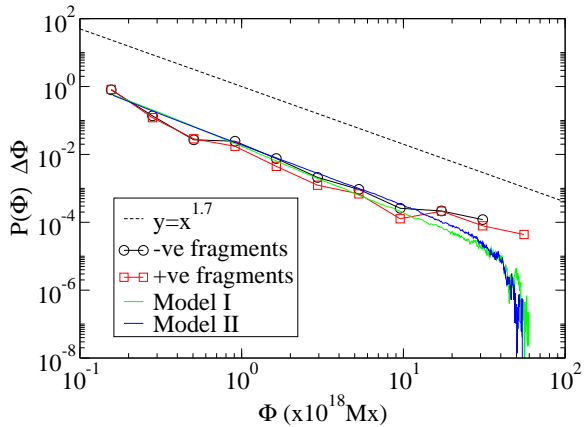} \caption{The degree distribution of the magnetic network. The
normalized number of magnetic concentrations in bins of size $\Delta \Phi =
1.55\times 10^{17}$ Mx obtained by reanalyzing the measurement data originally
shown in Figure 5 of \citet{close}. The straight line has a slope -1.7
corresponding to equation \ref{eq:phidphi}. The model data shown represents the
probability distribution, $P(k_{foot})$, for number of loops, $k_{foot}$,
connected to a footpoint. This has been rescaled so that one loop,
$k_{foot}=1$, equals the minimum threshold of flux, $1.55 \times 10^{17}$ Mx
(see equation \ref{eq:ktophi}). The parameters used were $m=0.1,L=100$ for
model I and $m=1,L=100$ for model II.} \label{fragmentconnections}
\end{figure}

The data was originally presented using a log-linear plot, which is appropriate
for distributions with one, or a few, characteristic scale(s), such as the
exponential distribution. We have reanalyzed this data to determine the
normalized histogram of concentration strengths, as shown in Figure
\ref{fragmentconnections}, plotted on a double logarithmic scale.

First, the fraction of positive (negative) magnetic fragments, or
concentrations, in bins of size $\Delta \Phi=1.55\times10^{17} Mx$ were
computed from the cumulative distributions reported in Figure 5 of
\citet{close}. This value of $\Delta \Phi$ was chosen to correspond to the
minimum threshold resolution reported in Close et al. For each bin, this
fraction, $P(\Phi) \Delta \Phi$, is the number of concentrations in that bin
divided by the total number of concentrations detected.  Then we aggregated
these bins into logarithmically increasing ones, and divided by the number of
$\Delta \Phi$ bins in each logarithmic one. The logarithmic aggregation is done
solely to increase the accuracy at larger fragment sizes.

As demonstrated in Figure \ref{fragmentconnections}, within statistical
uncertainty, the distribution of fragment sizes is a power law both for
positive and negative concentrations over the entire range of the measurement,
about $(2-500)\times 10^{17}$ Mx; i.e.
\begin{equation}
P(\Phi) \Delta \Phi \sim \left({\Phi\over \Delta \Phi}\right)^{-\gamma} {\rm \
with\ } \gamma=1.7 \pm 0.3 \, .\label{eq:phidphi}
\end{equation}
This shows that magnetic concentrations form a scale free network.

\subsection{Model Results and Comparisons}

The number of loops attached to a footpoint is an integer, $k_{foot}$, which is
greater than or equal to one.  The distribution $P(k_{foot})$ measures the
likelihood that a footpoint selected at random will have $k_{foot}$ loops
attached to it. This number corresponds to the amount of flux observed in a
concentration on the photosphere, since each concentration is identified by a
single footpoint locating its center.  In the context of networks, the integer
$k_{foot}$ is normally written simply as $k$, which is the number of links
attached to a node. The quantity $k$ is referred to as the degree of the node,
and $P(k)$ is referred to as the degree distribution of the network.

Figure \ref{fragmentconnections} compares results from numerical simulations of
models I and II with the observational data. It is necessary to calibrate units
of flux in the model to actual flux on the photosphere.  The  integer loop unit
in the model was set equal to the minimum threshold resolution used by Close et
al, i.e.
\begin{equation}
\Phi = k_{foot}\times(1.55\times10^{17}{\rm Mx}). \label{eq:ktophi}
\end{equation}
Otherwise, the comparison involves no fitting parameters. Over the range where
concentrations were detected, agreement between the reanalyzed observational
data and the model results are excellent. Equivalent to Eq.~\ref{eq:phidphi},
the degree distribution of the scale free magnetic network is
\begin{equation}P(k_{foot}) \sim k_{foot}^{-\gamma}\label{eq:pofk}\end{equation}
where the critical exponent $\gamma$ is the slope of the curves shown in Figure
\ref{fragmentconnections}.  Note that if the minimum threshold resolution were
changed, the calibration expressed in Eq.~\ref{eq:ktophi} would change
correspondingly.

\subsubsection{Universality classes}

\begin{figure}[t]
\plotone{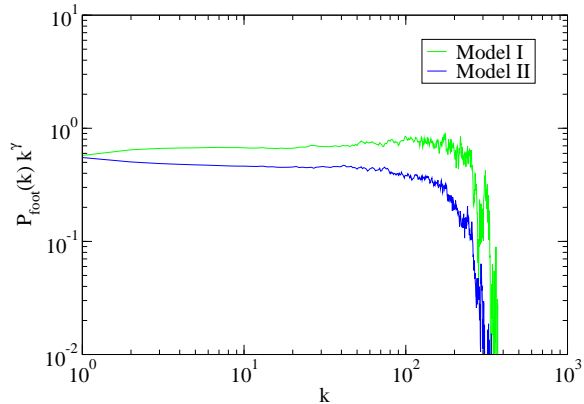} \caption{Scale free network exponent. The probability
distribution, $P(k_{foot})$, for number of loops, $k_{foot}$, connected to a
footpoint, multiplied by $k_{foot}^{\gamma}$. Models I and II require different
values of gamma in order to obtain a horizontal line over the central scaling
region. In this figure we use the values $\gamma=2.0$ for model I, and
$\gamma=1.75$ for model II.} \label{fragmentconnections-slope}
\end{figure}

The critical exponent for models I and II are further apart than Figure
\ref{fragmentconnections} may suggest. For model I, $\gamma=2\pm0.1$, whereas
for model II, $\gamma=1.75\pm0.1$. In Figure \ref{fragmentconnections-slope},
we have multiplied $P(k_{foot})$ by $k_{foot}^2$ for model I and by
$k_{foot}^{1.75}$ for model II. This tests for any systematic deviations from
scaling. The resulting horizontal lines over a region spanning more than two
decades provides a strong indication that the numerical simulation results are
probing an asymptotic scale-free network.  As the exponent $\gamma$ differs
between model I and II, concentration cancelation on the surface of the
photosphere appears to change the universality class of the scale-free magnetic
network.

\subsubsection{The range of scaling}

Note that the cutoff in the degree distributions shown in Figure
\ref{fragmentconnections} and \ref{fragmentconnections-slope} is a finite size
effect.  With respect to the numerical simulation results, the range of scaling
can be increased by e.g. increasing the system size ratio $L/l_{min}$. This
allows more footpoints and loops into the system.  Since the number of loops
attached to any footpoint cannot exceed the total number of loops in the
system, increasing $L/l_{min}$ extend the scaling range.

With respect to the observational data, the range of scaling can be extended at
both the low and high flux ends. Longer observation time would make it more
likely to observe large flux concentrations, which occur with relatively low
probability, extending the scaling regime to higher flux strengths.  A finer
flux resolution would allow smaller flux concentrations to be detected,
extending the range at low flux strengths.

\subsection{Previous Empirical Results}

Measurements of the distribution of magnetic concentrations on the quiet-Sun
have been made by several groups. \citet{schrijver_flux} analyzed high
resolution magnetograms from the MDI instrument on SoHO. They applied a number
of filtering and selection techniques to identify ephemeral regions. They found
an approximately exponential distribution of magnetic concentration strengths
between about $10^{18}$ and about $5 \times 10^{18}$ Mx, with a raised tail
above that range. They also proposed a model, incorporating flux emergence,
merging, cancelation and fragmentation, which has an exponential distribution
within a certain range.  \citet{hagenaar2001} also reported an exponential
distribution in the range $ (3 - 8.5) \times 10^{18}$ Mx.

Further observations were made by \citet{hagenaar2003}. Data was obtained for
different years, from 1996 to 2001 corresponding to roughly half a solar cycle.
They found that the distributions of magnetic concentration sizes measured at
different points in the solar cycle could be more accurately described by a sum
of two exponentials over the range $(3 - 100) \times 10^{18}$ Mx. Their data
analysis used a different algorithm to define concentrations than used by Close
et al, including smoothing the magnetogram data by convolution with a Gaussian
and the imposition of a filter to remove active regions.

\subsubsection{A reanalysis of Hagenaar et al data}

\begin{figure}[t]
\plotone{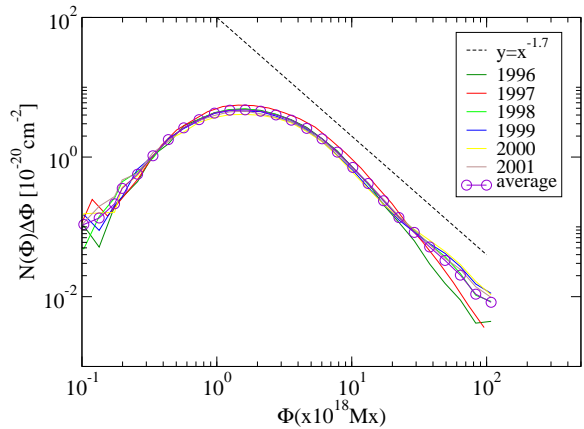} \caption{Replotted distribution of magnetic flux measured by
\citet{hagenaar2003}. Above $\sim 4\times10^{18}$Mx the distribution is well
approximated by a power law with exponent $\sim-1.7$. The roll-over at small
flux values may be due to the data acquisition procedure, and is not evident in
the reanalyzed data from \citet{close} shown in Figure
\ref{fragmentconnections}.} \label{hagenaar}
\end{figure}

In Figure \ref{hagenaar}, the observation data from Figure 5 of
\citet{hagenaar2003} has been replotted on a double logarithmic scale. As
shown, all six data sets, representing different phases of the solar cycle, are
consistent with a power law tail.  The figure displays the magnetic
concentration counts  for each year, along with an average over all six data
sets. The different years follow the same curve almost exactly below
$\sim30\times10^{18}$Mx, but differ somewhat above this value. A likely
explanation is statistical noise, although it is not possible to rule out a
systematic variation with the solar cycle.

It is evident that our model describes the data obtained by Close et al better
than the data obtained by Hagenaar et al. This is particularly apparent at
small concentration sizes.  However, the roll-over at small concentrations seen
in Figure \ref{hagenaar} may be an artifact of the smoothing algorithm used in
the data acquisition procedure used by Hagenaar et al and not by Close et al.
In fact, a roll-over is also found in numerical simulations of our model by
imposing a finite resolution grid, as described later.

\subsubsection{Active Regions}
Considering larger concentrations of magnetic flux, \citet{harvey:active}
report the distribution of active region areas.  The observed counts can be
described as an approximate power law with an index, $\gamma \simeq 2$, from
about three square degrees to about forty square degrees.  This index is very
close to the index shown in Figure \ref{fragmentconnections}, for smaller
magnetic concentrations on the quiet-Sun.  Thus there is some empirical
evidence to support our conjecture that the distribution of flux concentrations
is scale invariant over a range spanning the smallest (currently) measurable
concentrations to the large active regions.  In other words, there is one power
law for the distribution of all flux concentrations regardless of their
strength, up to a cut-off set by the size of the Sun.

In numerical simulations of our model, unsurprisingly, the largest avalanches
of reconnection emanate from flux tubes attached to the largest concentrations.
The big concentrations could therefore be referred to as `sunspots' or large
opposite polarity pairs referred to as `active regions'.  However, it is
generally believed that active regions have a different origin and dynamics
than smaller concentrations on the quiet-Sun. There are correlations in the
dynamics of flux emergence which are not included in our model. Such
correlation, if they were characterized and quantified, could be incorporated
into the driving term(s) - e.g. steps 3 and 4.

\subsection{Net Signed Flux in Grid Cells}

The scaling behavior of critical systems is robust to coarse-graining. In fact,
this is the fundamental concept behind the renormalization group, and scale
invariance in general.  This means that the system displays the same
statistical behavior regardless of the scale of observation.  Finite instrument
resolution does not alter the critical properties, but only the range of
observable scaling behavior. Such measurement imprecision is typically
encountered at small values, i.e. short length scales or low flux values. The
effect on the distribution is only seen at these low values, usually as a
flattening of the distribution near the resolution limit.

We illustrate the effect of finite resolution in the model by imposing a grid
with cells of linear extent $l_{cell}$ on the system and counting the net
signed flux associated with all footpoints in each cell. Since the individual
distribution of footpoint strengths has a Levy tail with an infinite variance
(Eq. \ref{eq:pofk}), this sum is not expected to converge to a Gaussian
according to the central limit theorem. Instead, the sum is dominated by the
largest instance in the sample, and converges to a Levy-stable distribution.
Thus the distribution of net signed flux within a grid cell asymptotically
behaves as
\begin{equation}
P\left(\; \left|\sum_{cell} k'_{foot} \right|\;\right) \sim
\left(\;\left|\sum_{cell} k'_{foot}\right|\;\right)^{-\gamma} \quad ,
\end{equation}
where $k'_{foot}=k_{foot}$ for positive footpoints and $k'_{foot}=-k_{foot}$
for negative footpoints.

\begin{figure}[t]
\plotone{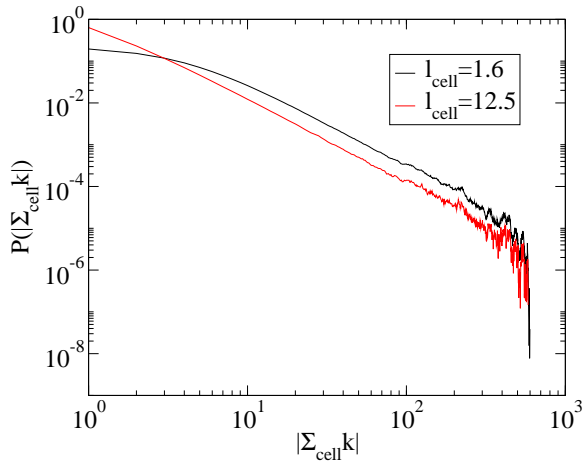} \caption{Net signed flux within grid cells of length
$l_{cell}$ for model I, with $L=100$. This shows a simulated effect of finite
resolution on measurements of concentration flux. Note that the power law is
robust and only the smallest scales are affected by resolution. We suspect that
allowing concentrations to occupy a finite area in the model (as on the
photosphere) would cause the cutoff at large scales to move to higher values
with increasing $l_{cell}$.} \label{grid}
\end{figure}

As shown in Figure \ref{grid}, coarser resolution causes a round-off at small
values of $ \left(\sum_{cell} k\right)$.  However, the scaling behavior of the
probability distribution of net signed flux within a grid cell is a power-law
with the same index as for the individual footpoints, and independent the size
of the grid cell, $l_{cell}$.  In the model, the cutoff at large scales occurs
at the same point for all grid cell lengths. It is determined by the global
system size $L/l_{min}$. This is due to the unrealistic assumption that the
footpoints, or magnetic concentrations, do not occupy any physical area on the
surface. Associating a physical area to the footpoints would likely cause the
large scale cutoff in the distribution to depend also on $l_{cell}$. The
largest footpoints could cover an area larger than an individual grid cell.

\subsubsection{A Grid Measurement}
The observation of robust power law behavior, independent of the grid cell size
over a range of scales, suggests a measurement of photospheric magnetic flux
that does not require the identification or selection of concentrations.  Nor
does it require filtering images to remove active regions or any other
features.  An unbiased measurement for the sun, corresponding to that described
here for the model, would involve imposing a mathematical grid with grid cell
length of a specified size $l_{cell}$, on magnetometer images of the
photosphere. The net signed magnetic flux in each grid cell $|\Phi_{cell}|$ is
calculated by summing all the signed values of flux $\phi$ in pixels within the
grid cell, therefore
\begin{equation}
\left|\Phi_{cell}\right| = \left|\sum_{cell}\phi\right|. \label{signedflux}
\end{equation}

A large statistical sample could be obtained by collecting magnetometer images
at different times.  For each $l_{cell}$, this allows a computation of the
probability distribution of net signed flux, $P(|\Phi_{cell}|)$.  The model
results indicate that this would yield a power law distribution of flux
concentrations
\begin{equation}
P(|\Phi_{cell}|)\Delta \Phi \sim{\left( |\Phi_{cell}| \over {\Delta
\Phi}\right)}^{-\gamma},
\end{equation}
for a range of $l_{cell}$.

The cutoffs at both small and large values of flux may depend on $l_{cell}$.
If the cutoff at large scales increases with $l_{cell}$, data collapse methods
could be used to rescale the flux according to the cell size, so that all the
distributions would coincide on the rescaled plot. We expect a similar behavior
to obtain by neglecting the sign of the flux, i.e. the probability distribution
for net unsigned flux within a grid cell, $P(\sum_{cell}|\phi|)$, may also be
scale-free.

\section{The Flux  Network}

The magnetic concentrations themselves do not comprise a network. They must be
joined by magnetic fields, or flux tubes, in order to do so. In fact
\citet{close} also measured statistical properties of the magnetic flux network
in the quiet-Sun that allow comparisons with our model.  They found that (1)
for any concentration strength there are a wide range of possible connections;
(2) concentrations show a preference towards connecting to nearby opposite
polarity concentrations; (3) despite the vast number of possible connections,
the bulk of the flux is often divided so that most of it goes to one opposite
polarity concentration.

Similar results are found in numerical simulations of our model.  In fact, all
of this behavior is explained by the existence of three different scale-free
distributions.  The first is  the distribution of concentration sizes (node
degree) as discussed previously.  The second and third are new statistical
quantities we introduce to characterize networks:  (a) the amount of flux
connecting a pair of concentrations, or the strength of the link between a pair
of nodes, and (b) the number of distinct concentrations (nodes) linked to a
given one. These are both found to be power laws in the model, with different
indices. We also discuss the distribution of the lengths of flux tubes and
compare with observations of Close et al. This allows a calibration of the
length unit in the model to a physical length on the photosphere.

\subsection{The strength of connections}

\begin{figure}[t]
\plotone{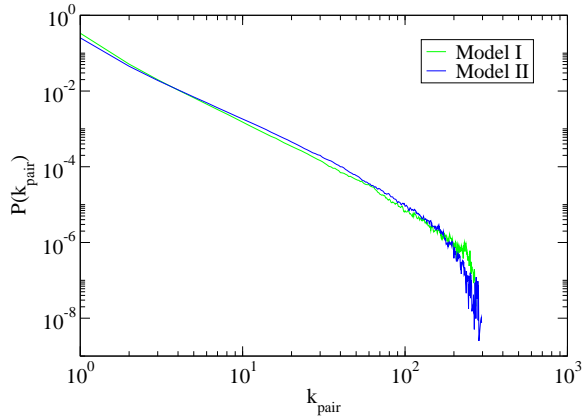} \caption{The link strength distribution. The probability
distribution, $P(k_{pair})$, of the number of loops, $k_{pair}$, connecting
pairs of footpoints. $P(k_{pair})$ follows a power law with exponent
$\alpha=2.25\pm0.1$ for model I and $\alpha=2.0\pm0.1$ for model II. The
parameters used were $m=0.1,L=100$ for model I and $m=1,L=100$ for model II. }
\label{kpair}
\end{figure}

Two opposite polarity footpoints can be connected to each other by any number
of loops. The number of loops connecting a pair of footpoints is defined as
$k_{pair}$. This is the strength of the link between two nodes. Measuring this
value over all footpoint pairs gives the distribution shown in Figure
\ref{kpair}. Power law behavior occurs for both models, i.e.
\begin{equation}
P(k_{pair}) \sim k_{pair}^{-\alpha}.
\end{equation}
The critical index $\alpha$ is equal to $2.25\pm0.1$ and $2\pm0.1$ for models I
and II respectively.  The index $\alpha\geq\gamma$, as the number of loops
connecting any pair of footpoints cannot exceed the number of loops at one of
those footpoints. The cut-off at large flux values in Figure \ref{kpair} is a
finite size effect; increasing number of loops in the system (for example by
increasing the system size) will extend the scaling region and move the cut-off
to higher values. Simulation results lead us to propose that the distribution
of amount of magnetic flux connecting any pair of concentrations on the
photosphere is distributed as a power law, with an index $\alpha$.

\subsection{The number of unique connections}

\begin{figure}[t]
\plotone{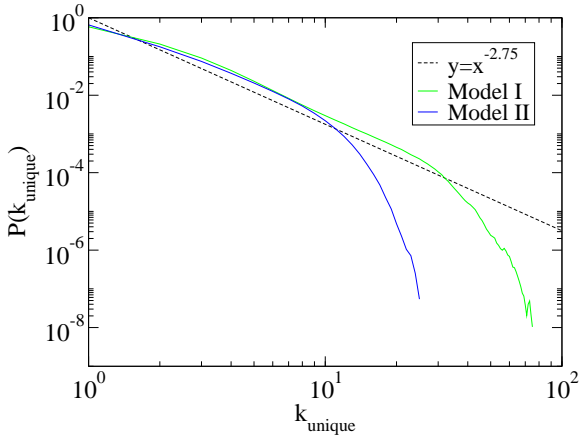} \caption{The unique connection distribution, $P(k_{unique})$,
for the number of unique footpoints to which a given footpoint is linked. The
straight line indicates a power law with exponent $\beta\sim2.8$. The
parameters are $m=0.1$, $L=100$ for model I, and $m=1$, $L=100$ for model II.}
\label{kunique}
\end{figure}

One footpoint can be connected to an arbitrary number of opposite polarity
footpoints. For each footpoint we count the number of distinct footpoints that
are connected to it. This is  $k_{unique}$. Multiple loops between a pair of
footpoints are counted as a single connection in this measurement. This
distribution, shown in Figure \ref{kunique}, may be consistent with power law
behavior
$$P(k_{unique}) \sim k_{unique}^{-\beta}.$$
The exponent $\beta\sim2.8$ for model I.  Since $\beta$, if it exists, is
greater than 2, the average number of unique footpoints connected to any given
one is finite. However, if $\beta \leq 3$ then the variance in the number of
unique connections diverges.

The quantities $P(k_{pair})$ and $P(k_{unique})$ are complementary measures of
the network structure. The former measures the strength of connections, the
latter measures the number of unique connections. Since the average number of
unique connections is probably finite, concentrations with high flux values are
more likely to have a small number of strong connections than a large number of
weak ones. These results agree qualitatively with observations of Close et al
listed at the start of section 4. It may be possible to measure the quantities
$P(k_{pair})$ and $P(k_{unique})$ on the Sun.

It is not required that both $P(k_{pair})$ and $P(k_{unique})$ follow power law
behavior in order to obtain a scale free network. The only requirement is that
the degree distribution $P(k_{foot})$ is scale free. For instance, in the case
of the World Wide Web \citep{barabasi99}, the scientific citation network
\citep{redner:citations} and most, if not all, previously studied examples
\citep{bornholdt,barabasi02} the link strength has been defined to be either
one or zero. For those networks $\beta$ must be equal to $\gamma$. One can
imagine the extreme opposite case. Then the number of distinct nodes connected
to any give node would not exhibit power law behavior. It appears that the
model operates in between these two extremes, as a fully scale free network,
where all three distributions are power laws.

Many scale free networks also display a higher clustering coefficient than
corresponding random networks \citep{maslov02}. The clustering coefficient
measures the fraction of nodes that are linked to a given node and also to each
other. Graphically, this is the number of triangles in the network. In the
coronal magnetic network these triangles cannot occur, as same polarity
footpoints cannot link directly to each other. However the number of
quadrilaterals in the network can be measured. A quadrilateral is formed when
two footpoints (nodes) of the same polarity both have links to the same two
opposite polarity footpoints. It is possible that the clustering coefficient
for our model, and for the coronal magnetic field,  measured in this way would
be higher than in a randomized network.

Another important property of scale free networks is that the average distance
between nodes is low, i.e. it is possible to reach a given node from any other
visiting a small number of intermediate nodes. A short average distance is
observed in both scale free and random networks, but not in ordered networks
(such as lattices or strictly hierarchical networks).  It is possible to
measure this quantity for our model as well as the coronal magnetic field.

\subsection{Flux Tubes}

\begin{figure}[t]
\plotone{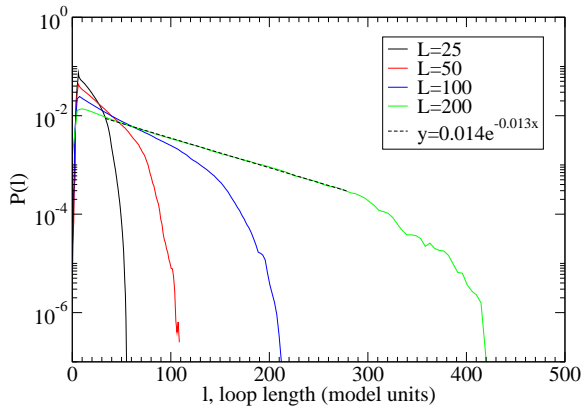} \caption{The probability distribution, $P(l)$, of loop
lengths, $l$. $P(l)$ exhibits exponential behavior for each system size $L$
over the range $0.05\pi L<l<0.45\pi L$. A finite size cutoff occurs where
boundary effects limit the maximum loop length. Results are shown for model I
with $m=0.1$.} \label{lengths}
\end{figure}

For each pair of connected footpoints, we measured the length of the loop
linking them, irrespective of the strength of the link. This was done according
to the definition of loop length in section 2.2. The distribution of loop
lengths measured in numerical simulations is shown in Figure \ref{lengths}. The
loop lengths appear to be exponentially distributed from approximately $0.05\pi
L<l<0.45\pi L$, i.e. $$ P(l)\sim e^{-l/\xi(m,L)}$$ where $\xi(m,L)$ is the
characteristic length of a loop. In Table \ref{varyL}, $\xi$ is determined by
fitting to the data for each system size between $0.05\pi L$ and $0.45 \pi L$
using least squares analysis. The fit to an exponential in this range is
extremely good (e.g. $r>0.997$ for $L=200$). Occasionally loops extend up to
the maximum possible size $\pi L / \sqrt{2}$. Increasing the system size, $L$,
not only increases the maximum observed loop length, but also changes $\xi$. As
shown in table \label{varyL}, there is a weak dependence of $\xi$ on the
stirring rate $m$ for a fixed system size $L$.

As flux tubes in our model are described as semi-circular loops normal to the
photosphere, the footpoint separation and maximum loop height are also
exponential distributions, trivially related to $P(l)$.

\begin{figure}[t]
\plotone{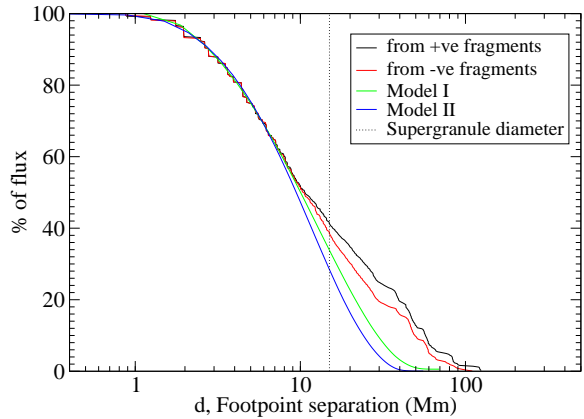} \caption{The cumulative percentage of pairs of connected
concentrations separated by a distance larger than $d$. The flux tube data
corresponds to Figure 6c in \citet{close}. The model data has been scaled such
that one unit of length is equal to 0.5Mm. Good agreement is found for
separations up to the typical supergranule diameter, 15Mm. The parameters used
were $m=0.1,L=100$ for model I and $m=1,L=100$ for model II.}
\label{separationcompare}
\end{figure}

\begin{figure}[t]
\plotone{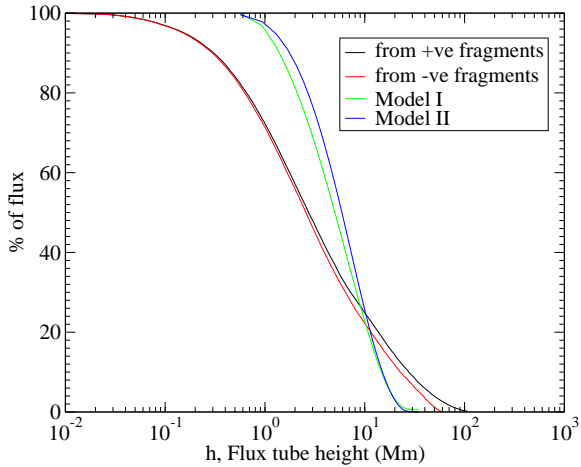} \caption{The cumulative percentage of maximum loop heights
larger than a certain amount. The flux tube data corresponds to Figure 6b from
\citet{close}. The model data has been scaled such that one unit of length is
equal to 0.5Mm, as in Figure \ref{separationcompare}. Agreement between the two
data sets is poor, and the range of heights obtained by \citet{close} under the
potential field approximation is larger than is possible with our model with
the given parameters. The parameters used were $m=0.1,L=100$ for model I and
$m=1,L=100$ for model II.} \label{heightcompare}
\end{figure}

Close et al calculated flux tube statistics from magnetograms based on a
potential field approximation. Figures \ref{separationcompare} and
\ref{heightcompare} compare their results for footpoint separation and maximum
height of flux tubes with Models I and II. Note that both of these figures show
the cumulative distributions.

Reanalysis of the footpoint separation data from Close et al reveals that the
probability distribution of footpoint separations follows an exponential with a
characteristic length of $20\pm3$Mm.

\subsubsection{Calibration}

Figure \ref{separationcompare} demonstrates how the unit length of the model
can be calibrated to actual lengths on the photosphere. Comparing the
cumulative distribution of the model to Close et al, and setting one unit of
length in the model equal to 0.5Mm for $L=100$, the cumulative distribution of
footpoint separations agree up to the supergranule cell size, (15Mm). Beyond
that length scale, the distributions do not agree. This difference may be due
to the finite size effects in the model and/or the potential field
approximation used to determine the flux tubes.

\subsubsection{Flux tube heights}

Figure \ref{heightcompare} shows the cumulative distribution of maximum loop
heights for model I and II compared to the flux tube heights measured by Close
et al. Using the same calibration factor as determined for lengths in Figure
\ref{lengths}, the results from the numerical simulations do not agree with the
measurements over the entire range. Although the distributions have roughly the
same shape, the heights calculated by Close et al extend over a much wider
range than is seen or is indeed possible in the model discussed here. This
disagreement is not particularly surprising. The semi-circular loops in our
model are always perpendicular to the photosphere, so that the maximum height
attained is simply the radius of the loop. Under the potential field
approximation, the maximum height a loop can attain is not simply related to
the footpoint separation. For example the loop can trace out a very flat curve,
lying close to the photosphere. However, TRACE images tend to qualitatively
support the picture of flux tubes emerging perpendicular to the photosphere.

\subsection{Emergence rate and turnover time}

\begin{table}[t]
\begin{tabular}{@{\extracolsep{\fill}}|c|c|c|c|}
\hline
m & $\xi$ (model units) & $\langle N\rangle$ & $\langle N_{loops} \rangle$\\
\hline
0.001 & 22 & 549 & 1100 \\
0.01 & 31 & 358 & 720 \\
0.1 & 40 & 190 & 391 \\
1 & 43 & 83 & 139 \\
\hline
\end{tabular}
\caption{Average number of footpoints $\langle N\rangle$, average number of
loops $\langle N_{loops} \rangle$ and characteristic loop length $\xi$ in model
I for different values of $m$, with system size $L=100$. Note that $\xi$
depends very weakly on $m$.}\label{varym}
\end{table}

In order to determine the predicted flux turnover time for the Sun we need to
calibrate not only the units of distance but also of time in the model. The
length calibration was determined earlier to be one unit of length equal to
0.5Mm for Model I with $m=0.1, L=100$. Strictly this calibration varies for
different values of $m$. However, as can be seen in table \ref{varym} the
characteristic loop length $\xi$ only changes by a factor of about 2, while $m$
is varied over 3 orders of magnitude. Therefore the effect of changing $m$ on
the length scale is very small and is ignored in the calculation that follows.

\citet{hagenaar99} observe that magnetic concentrations diffuse on the
photosphere. Footpoints in our model also diffuse. They execute a random walk
with the distance moved in each discrete time step drawn from a uniform
probability distribution between 0 and 1, or 0 to 0.5Mm after calibration for
$L=100$. The mean square displacement of the footpoints in one step is
therefore 0.083Mm$^2$. During a parallel update step each footpoint moves once
on average, therefore a parallel update step is equal to N single footpoint
updates, where N is the total number of footpoints in the system. We denote the
length of time taken for a parallel update step as $\Delta t$. From Figure 2 in
\citet{hagenaar99}, the mean square displacement of 0.083Mm$^2$ is observed to
take approximately 300 seconds. We therefore set $\Delta t=300$secs$=0.083$hrs.

The parameter $m$ measures the fraction of footpoints moved between loop
injections, therefore in one parallel update step $1/m$ loops are added. The
number of loops injected per hour is given by
\begin{equation}
loops/hour = \frac{1}{m\Delta t}, \label{loopsperhour}
\end{equation}
where $\Delta t$ is measured in hours.

We determine the turnover time, $t_{turnover}$, from the average number of
loops in the system, $\langle N_{loops} \rangle$, and the number of loops
injected per hour by the following formula:
\begin{equation} t_{turnover} = \frac{\langle N_{loops}\rangle}{loops/hour} = m\Delta t \langle N_{loops}\rangle .
\label{turnover}
\end{equation}
The turnover times for model I and II are shown in tables \ref{sunflux-1}, and
\ref{sunflux-2} respectively. Note that the flux calibration used in Figure
\ref{fragmentconnections} is not relevant to this calculation.

As seen in tables \ref{sunflux-1} and \ref{sunflux-2} and expressed in equation
\ref{turnover}, the turnover time depends on the stirring rate $m$. However in
model I, which includes the cancelation of concentrations on the photosphere,
the turnover time has a much stronger dependence on the stirring rate than it
does in model II, which does not allow the cancelation of concentrations. At
this point in time we have no a priori reason to select model I or model II or
a particular stirring rate, so we cannot make a completely definitive statement
as to the expected turnover time.

However it is reasonable to expect that the dimensionless parameter $m$, that
characterizes how the flux is driven in the system, should be neither extremely
large or extremely small. If $m$ is very large then essentially all the flux is
removed from the system and the density of footpoints is vanishingly small.
While if $m$ is very small then the footpoints barely move and cannot be
observed to diffuse during their lifetime. The fact that we see flux emerging
as well as the diffusion of concentrations indicates that $m$ should be
approximately within the range of values shown in tables \ref{sunflux-1} and
\ref{sunflux-2}.

\subsubsection{Emergence rate}

Tables \ref{sunflux-1} and \ref{sunflux-2} also show the flux emergence rate
and total amount of flux for the entire solar surface estimated from the
numerical simulation results as explained below. Unlike the turnover time,
these estimates require the calibration of flux in the model to measurable flux
on the photosphere. We use the same calibration as in Figure
\ref{fragmentconnections}. One unit of flux is equal to $1.55\times10^{17}$Mx,
which is the threshold resolution of Close et al. The results from the
simulations were scaled up to the entire solar area by assuming the system of
size $L$ represents a typical section of the photosphere. Thus the flux values
are multiplied by the ratio of the total photospheric surface area to the model
surface area (50Mm$\times$50Mm for $L=100$).

\begin{table}[t]
\begin{tabular}{@{\extracolsep{\fill}}|c|c|c|c|}
\hline
m & emergence & total flux & $t_{turnover}$\\
&rate (Mx/hr) &(Mx) &(hr)\\
 \hline
0.001 & $5\times10^{24}$ & $5\times10^{23}$ & 0.1\\
0.01 & $5\times10^{23}$ & $3\times10^{23}$ & 0.6\\
0.1 & $5\times10^{22}$ & $2\times10^{23}$ & 3\\
1 & $5\times10^{21}$ & $6\times10^{22}$ &  12\\
\hline
\end{tabular}
\caption{Average flux emergence rate, total amount of flux, and flux turnover
time in model I with $L=100$, scaled to represent the total surface area of the
Sun. }\label{sunflux-1}
\end{table}

\begin{table}[t]
\begin{tabular}{@{\extracolsep{\fill}}|c|c|c|c|}
\hline
m & emergence  & total flux& $t_{turnover}$\\
 & rate (Mx/hr) & (Mx) & (hr)\\
\hline
0.1 & $5\times10^{22}$ & $6\times10^{23}$ & 12\\
1 & $5\times10^{21}$ & $1\times10^{23}$ & 25\\
10 & $5\times10^{20}$ & $3\times10^{22}$ & 63\\
\hline
\end{tabular}
\caption{Average flux emergence rate, total amount of flux, and flux turnover
time in model II with L=100, scaled to represent the total surface area of the
Sun.} \label{sunflux-2}
\end{table}

It is immediately clear that the total flux does not vary linearly with $m$,
although by equation \ref{loopsperhour} the rate of flux emergence does. This
means that the model can exhibit a range of flux turnover times on changing
$m$. This agrees with the physical picture where higher injection rates (when
$m$ is small) result in systems with a higher loop and footpoint density,
making interactions more frequent. In particular, concentration cancelation in
model I will occur more often (relative to parallel update steps) and hence
loop lifetimes will be reduced and flux turnover times decreased. This explains
why model I has a stronger dependence of turnover time on the stirring rate
$m$.

\subsubsection{Comparison with observations}

Studies measuring the rate of flux emergence have determined rates which vary
widely, from e.g. $\sim4\times10^{20}$Mx/hr \citep{harvey1993}, to
$\sim2\times10^{22}$Mx/hr \citep{hagenaar2001}. Later studies have tended to
indicate a higher rate of flux emergence, as ever smaller flux concentrations
are resolved and included in the analysis. The estimates of total solar flux
have hardly changed over the same time period, usually quoted as being no less
than $\sim3\times10^{23}$Mx (e.g. \citet{schrijver_flux}). The figures for flux
turnover rates have therefore correspondingly decreased over time. The recent
measurements by \citet{hagenaar2003} suggest a turnover time of 8-19 hours,
much shorter than the 40-70 hours from a few years earlier
\citep{schrijver_flux}.

Both model I and II give total flux values comparable with the estimates of
total solar flux, and both also give turnover times which are compatible with
current estimates. However it is necessary that one uses the same $m$ value for
both the total flux and the turnover time. This suggests that, based on current
estimates of the emergence rate and total flux, model II is a better
description of the coronal magnetic network than model I. Model II with
$m\simeq0.3$ agrees quantitatively with the current estimates for both the
total flux and the flux turnover time.

Thus the cancelation of concentrations may be a less important process in the
dynamics.

\section{Conclusions}

The main results of this work are as follows:
\begin{itemize}
\item Observational data of magnetic concentration strengths on the photosphere
were reanalyzed. We found that the distribution of concentration strengths
exhibits power law behavior, with a critical exponent $\gamma\sim1.7$, in the
range $2-500\times10^{17}$Mx. This distribution is compatible with that of
active region areas, which have much larger flux values.

\item We conjecture that a single power law describes the distribution of
magnetic concentrations of all sizes, from the smallest detectable fragments to
large active regions. The power law distribution at sufficiently large scales
is cutoff by the finite size of the Sun, as is typical in critical finite size
systems.

\item We propose that the coronal magnetic field embodies a scale free network.
Concentrations on the photosphere are nodes in the network. They are linked by
flux tubes. The distribution of magnetic concentration strengths therefore
corresponds to the degree distribution of a network, and it is scale free with
an exponent $\gamma$.

\item A self organized critical model involving avalanches of reconnecting flux
tubes generates a scale free magnetic network. The distribution of
concentration strengths measured in numerical simulations of the model agrees
quantitatively with the reanalyzed observational data presented here. A single
calibration is required that sets the unit of flux in the model equal to the
threshold magnetic flux that can be presently detected on the photosphere.
Otherwise there is no fitting parameter.

\item This model exhibits a power law distribution of flare events
\citep{modelprl}. Thus the scale free behavior of flares and of magnetic
concentrations are unified in terms of a single dynamical process, which self
generates the complex magnetic field structure and dynamics.

\item Two new quantities are introduced to characterize scale free networks.
Both of these quantities can be measured for the corona. The strength of a
connection between two nodes can be distributed as a power law, with an
exponent $\alpha\geq\gamma$. In the model this distribution is found to be
scale free. We propose that the amount of flux connecting two concentrations on
the photosphere has power law behavior with an exponent $\alpha$. Also the
number of unique concentrations linked to any concentration by flux tubes also
has a scale free distribution, with a different exponent.

\item The distribution of flux tube lengths measured in numerical simulations
of the model and observed in the corona is exponential up to the supergranule
cell size. This allows a calibration of the length unit of the model to a
physical length on the photosphere. Furthermore the observed diffusive behavior
of concentrations on the photosphere allows us to then calibrate the time unit
of the model.

\item Based on these calibrations we determine the flux turnover time and the
total coronal flux that the model predicts. Our results are compatible with
current estimated values for the Sun.

\end{itemize}

We thank R. Close and H.J. Hagenaar for generously sending us their previously
published observational data. We would like to thank R. Dendy and S.C. Cowley
for their comments, and P. Hanlon for his assistance with data processing. D.
Hughes acknowledges funding from EPSRC.

\bibliographystyle{apj}
\bibliography{ms}

\end{document}